# Anomalous relaxation kinetics and charge density wave correlations in underdoped BaPb$_{1-x}$Bi$_x$O$_3$

D. Nicoletti[1,*], E. Casandruc[1], D. Fu[1], P. Giraldo-Gallo[3,4],
I. R. Fisher[3,5], and A. Cavalleri[1,2]

[1] *Max Planck Institute for the Structure and Dynamics of Matter, 22761 Hamburg, Germany*
[2] *Department of Physics, Clarendon Laboratory, University of Oxford, Oxford OX1 3PU, United Kingdom*
[3] *Geballe Laboratory for Advanced Materials, Stanford University, Stanford, California 94305, USA*
[4] *Department of Physics, Stanford University, Stanford, California 94305, USA*
[5] *Department of Applied Physics, Stanford University, Stanford, California 94305, USA*
\* *e-mail: daniele.nicoletti@mpsd.mpg.de*

**Superconductivity often emerges in proximity of other symmetry-breaking ground states, such as antiferromagnetism or charge-density-wave (CDW) order. However, the subtle inter-relation of these phases remains poorly understood, and in some cases even the existence of short-range correlations for superconducting compositions is uncertain. In such circumstances, ultrafast experiments can provide new insights, by tracking the relaxation kinetics following excitation at frequencies related to the broken symmetry state. Here, we investigate the transient terahertz conductivity of BaPb$_{1-x}$Bi$_x$O$_3$ - a material for which superconductivity is 'adjacent' to a competing CDW phase - after optical excitation tuned to the CDW absorption band. In insulating BaBiO$_3$ we observed an increase in conductivity and a subsequent relaxation, which are consistent with quasiparticles injection across a rigid semiconducting gap. In the doped compound BaPb$_{0.72}$Bi$_{0.28}$O$_3$ (superconducting below T$_C$ = 7 K), a similar response was also found immediately above T$_C$. This observation evidences the presence of a robust gap up to T $\simeq$ 40 K, which is presumably associated with short-range CDW correlations. A qualitatively different behaviour was observed in the same material for T $\gtrsim$ 40 K. Here, the photo-conductivity was dominated by an enhancement in carrier mobility at constant density, suggestive of melting of the CDW correlations rather than excitation across an optical gap. The relaxation displayed a temperature dependent, Arrhenius-like kinetics, suggestive of the crossing of a free-energy barrier between two phases. These results support the existence of short-range CDW correlations above T$_C$ in underdoped BaPb$_{1-x}$Bi$_x$O$_3$, and provide new information on the dynamical interplay between superconductivity and charge order.**

The interplay between superconductivity and other broken symmetry ground states has emerged as a central theme in the physics of high-$T_C$ superconductors. For the Fe-based superconductors, the presence of phase transitions to states characterized by broken rotational symmetry and stripe-like magnetic order was established shortly after the initial discovery of superconductivity, though the relative role played by associated nematic and magnetic fluctuations with respect to the superconductivity remains an open question[1]. For the cuprates, although stripe order was recognized many years ago in several materials for doping close to 1/8[2], the existence of more subtle CDW order in other materials has only recently been established[3,4,5,6]. The inter-relation of these phases with superconductivity – whether they compete, coexist, or even whether associated fluctuations promote high critical temperatures – is an open question. These issues are not restricted to just Fe-based and cuprate superconductors, and important insights to the interplay of superconductivity and charge order can be found in other systems.

Although pressure[7] and magnetic fields[8] have long been used to affect the interplay between various broken symmetry states and superconductivity at low temperatures, only recently it has been shown that ultrashort optical pulses can achieve qualitatively similar effects over larger temperature ranges[9]. For example, coherent interlayer coupling could be induced in $La_{1.8-x}Eu_{0.2}Sr_xCuO_4$[10,11] and $La_{2-x}Ba_xCuO_4$[12,13,14] by optical melting of the stripe order[15]. In addition, optical excitation of apical oxygen distortions in $YBa_2Cu_3O_{6+x}$ was shown to induce a transient state far above $T_C$ with important similarities to the equilibrium superconductor[16,17,18]. This state is also characterized by a partial removal of CDW correlations[19].

Here we set out to explore the optical control of CDW states in a different family of materials, which have been dubbed "the other high-temperature superconductors"[20]: those based on doped $BaBiO_3$[21,22,23].

The bismuthates $BaPb_{1-x}Bi_xO_3$ (BPBO) and $Ba_{1-x}K_xBiO_3$ (BKBO) are examples of materials for which superconductivity is 'adjacent' to a competing charge ordered phase[24,25]. The 'parent' compound $BaBiO_3$ is insulating, comprising two distinct Bi sites to which formal valences can be assigned corresponding to 3+ and 5+. The static charge disproportionation (a so-called charge-disproportionated CDW state)[26] is associated with a frozen breathing distortion of the Bi-O octahedra, accompanied by subtle rigid rotations[27] (see crystal structure in the inset of Fig.1*A*).

When Ba is substituted with K[27], or Bi is substituted with Pb[28], the long-range charge order is rapidly lost, although these doped samples remain non-metallic up to relatively high doping levels. Indeed, the long range CDW that causes the insulating behavior in $BaBiO_3$, as well as the associated structural distortion, seem to persist on short length scales at intermediate concentrations, as inferred from Raman spectroscopy[29] and EXAFS measurements[30].

Doping with larger concentrations of Pb and K eventually leads to superconductivity, with maximum transition temperatures of 11 K in BPBO and 34 K in BKBO; values that are surprisingly high for such bad metallic oxides. Here, superconductivity is manifestly not due to spin fluctuations (none of the constituents show any tendency towards magnetism) and is more likely to arise from electron-phonon interactions.

Although the standard BCS theory applied to early determinations of the band structure failed to explain the relatively high $T_C$ values[31], recent calculations using more advanced functionals indicate that electron-phonon coupling in these materials is significantly enhanced by dynamic correlation effects[32]. While these calculations indicate that electron correlation effects should be included in order to accurately describe the electronic structure of the bismuthates, they are nevertheless based on the presumption of a homogeneous material, and neglect effects associated with short range electronic phase separation and/or fluctuations, leaving open questions associated with the role, if any, that competing CDW phases might play in determining the observed high critical temperatures.

The situation is further complicated by the presence of structural dimorphism in BPBO for superconducting compositions[33,34]. Such dimorphism is the result of thermodynamic equilibrium between two structural phases (tetragonal and orthorhombic), which are separated on a nanoscale. Importantly, the superconducting volume fraction has been found to scale with the relative proportion of the tetragonal phase, suggesting that only the tetragonal polymorph harbours superconductivity[33]. It is unclear what role this dimorphism might play in determining $T_C$ of the resulting matrix of interpenetrating polymorphs[34]. However, tunnelling measurements showed significant suppression of $T_C$ due to disorder[35], which implies that higher $T_C$ values might be obtained if disorder associated with structural dimorphism and chemical substitution could be somehow mitigated.

Ultrafast optical techniques provide a powerful new approach to understand and tune the bismuthate superconductors. Here, we take a first step in this direction, using pump-probe spectroscopy to uncover evidence for weak short-range CDW correlations with anomalous relaxation kinetics immediately above $T_C$ in underdoped BPBO.

We studied the terahertz-frequency (THz) photo-conductivity of the CDW insulator $BaBiO_3$, as well as of three superconducting compounds: $x = 0.28$ ($T_C = 7$ K), $x_{opt} = 0.25$ ($T_C = 11$ K), and $x = 0.20$ ($T_C = 7$ K) (see phase diagram in Fig. 1*A*), after femtosecond-pulse optical excitation tuned to the CDW absorption band. The single crystals were grown by slowly cooling a solution of $BaCO_3$, PbO, and $Bi_2O_3$. The bismuth concentration was determined by electron microprobe measurements, revealing a homogeneous composition (with an uncertainty of ±0.02) within a sample as well as for different samples from the same batch[36]. The structural dimorphism was previously characterized for samples grown under similar conditions[34]. For the compositions studied here, this dimorphism is established at very high temperatures, well above room temperature, providing a rigid framework in which CDW correlations and superconductivity evolve at much lower temperatures.

The equilibrium optical response of the compounds studied here was determined as a function of temperature with a Fourier Transform Infrared Spectrometer (Bruker Vertex 88v). The absolute reflectivity was measured in quasi-normal-incidence geometry in the ~1 – 150 THz range with the gold evaporation technique. By using literature data[37,38] for the higher frequency range, we could then perform Kramers-Kronig transformations and retrieve the complex optical conductivity.

Representative spectra of the material at equilibrium are displayed in Fig. 1*B*. These are in very good agreement with previous experiments[37,38], and show a fully gapped CDW insulator $BaBiO_3$, with at least four infrared-active phonon modes well visible in the 3 – 15 THz range. The first electronic absorption (*i.e.*, the CDW band, shaded in blue) starts to appear around 50 THz and peaks at ~350 THz (corresponding to 1.5 eV). Upon substitution of Bi with Pb, a narrow[39] Drude peak appeared below ~20 THz (green curve in Fig. 1*B*). At least in the *x* =0.28 sample, an absorption band centered around 50 THz (shaded in green), reminiscent of the CDW band, was still clearly visible. This observation indicates that CDW correlations may still be present locally at this doping level. We return to provide further evidence for this shortly, when we consider the transient response. When cooling below T$_C$, a conductivity gap is expected to open below ~1 THz[40]. We do not discuss these features here, and we will be focusing only on the normal-state response.

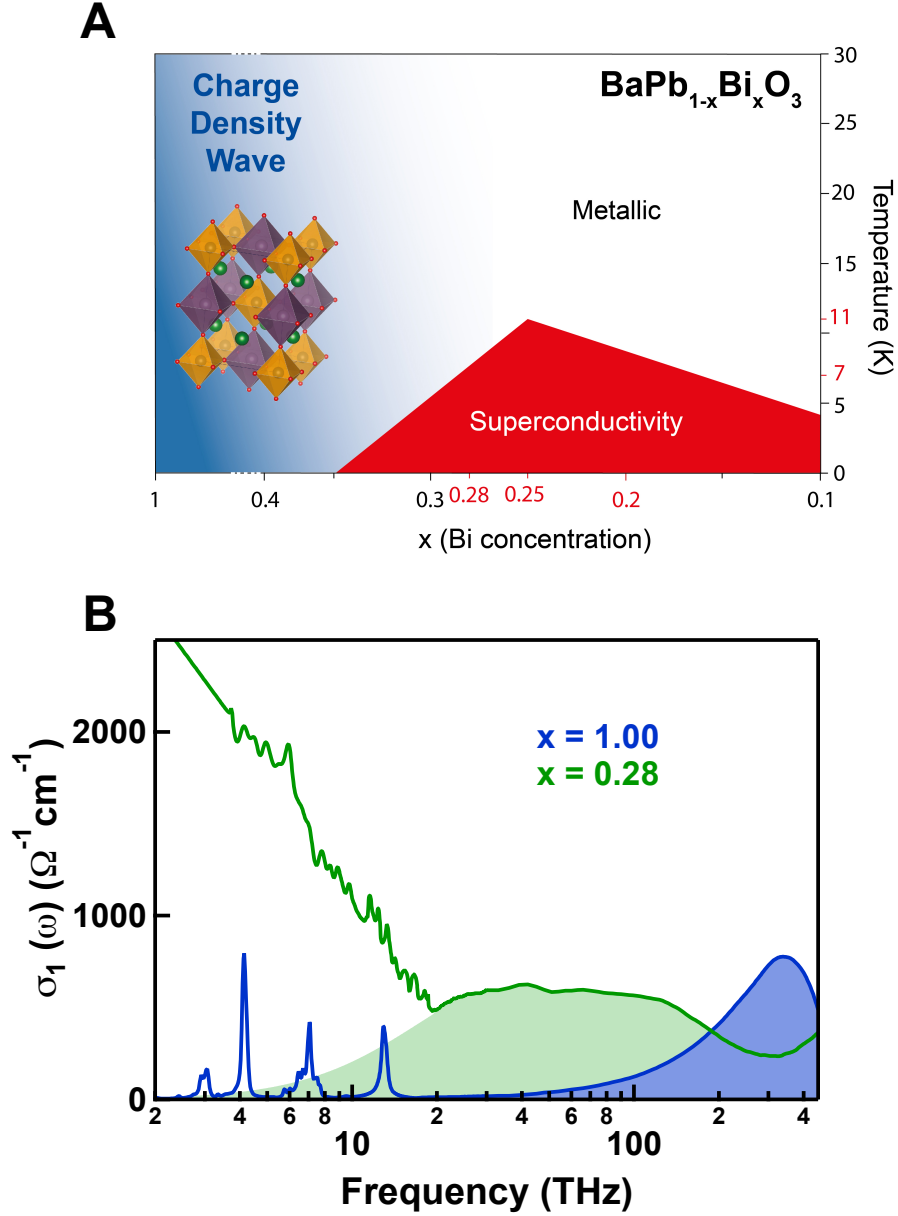

**Figure 1.** (*A*) Schematic temperature-doping phase diagram of BaPb$_{1-x}$Bi$_x$O$_3$, as determined in Ref. 45. The CDW and superconducting phases are shaded in blue and red, respectively. The Bi concentrations of the doped samples investigated in the present work, as well as their corresponding T$_c$'s, are highlighted in red. The crystal structure of BaBiO$_3$ is also displayed, comprising two distinct Bi sites, surrounded by oxygen octahedra with different Bi-O bond lengths (shown in yellow and violet, while green circles are Ba ions). (*B*) Equilibrium optical conductivity of BaBiO$_3$ and BaPb$_{0.72}$Bi$_{0.28}$O$_3$ measured at T = 15 K. The CDW band is highlighted in both spectra. Due to the quasi-cubic crystal structure of BPBO, no polarization dependence was found in any data reported in this paper.

In a series of pump probe experiments, each crystal was photo-excited with intense ~100-250 fs long laser pulses, whose frequency was tuned to be resonant with the CDW band (~375 THz in $BaBiO_3$, ~60 THz in the doped compounds, see Fig. 2). These pulses were generated either directly by a Ti:Sa amplifier (800 nm ~ 375 THz) or by converting the output of the laser to longer wavelengths with an optical parametric amplifier (5 μm ~ 60 THz). They were then focused to a ~2 mm spot onto the sample surface, at a fluence of ~3.5 mJ/cm². At these fluences, the excitation density across the CDW gap is of the order of $10^{21}$ carriers/cm³.

The transient photo-conductivity was then probed with delayed, quasi-single-cycle THz pulses, which were generated from a photo-conductive antenna and measured after reflection by electro-optic sampling in a ZnTe crystal. These measurements provided time and frequency dependent measurement of the complex optical properties between 0.5 – 2.5 THz[17,18].

The pump-induced changes in the amplitude and phase of the reflected THz electric field were measured for each sample at different pump-probe delays, τ. These "raw" reflectivity changes were only 1-3% percent, due to the pump-probe penetration depth mismatch. At THz frequencies, the probe interrogates a volume that is between 5 and 1000 times larger than the region beneath the surface transformed by the pump, with this mismatch being a function of frequency. Hence, the optical response induced by the pump pulse appeared smaller than what would have resulted from a homogenously excited sample. This mismatch was taken into account by modeling the response of the system as that of a photo-excited thin layer on top of an unperturbed bulk (which retains the optical properties of the sample at equilibrium). By calculating the coupled Fresnel equations of a multi-layer structure[41], the transient optical response (reflectivity, complex optical conductivity) of the photo-excited layer could be derived[16,17,18].

In figure 2 we show representative complex conductivities of three different sample measured at T = 15 K at equilibrium and after photo-excitation, at the peak of the pump-induced response (τ = 1 ps).

Resonant excitation at the CDW band in insulating $BaBiO_3$ (Fig. 2*A*) resulted in a prompt and complete filling of the conductivity gap, indicative of the generation of free carriers. Correspondingly, the quasi-DC conductivity reached values as high as ~100 $\Omega^{-1}cm^{-1}$.

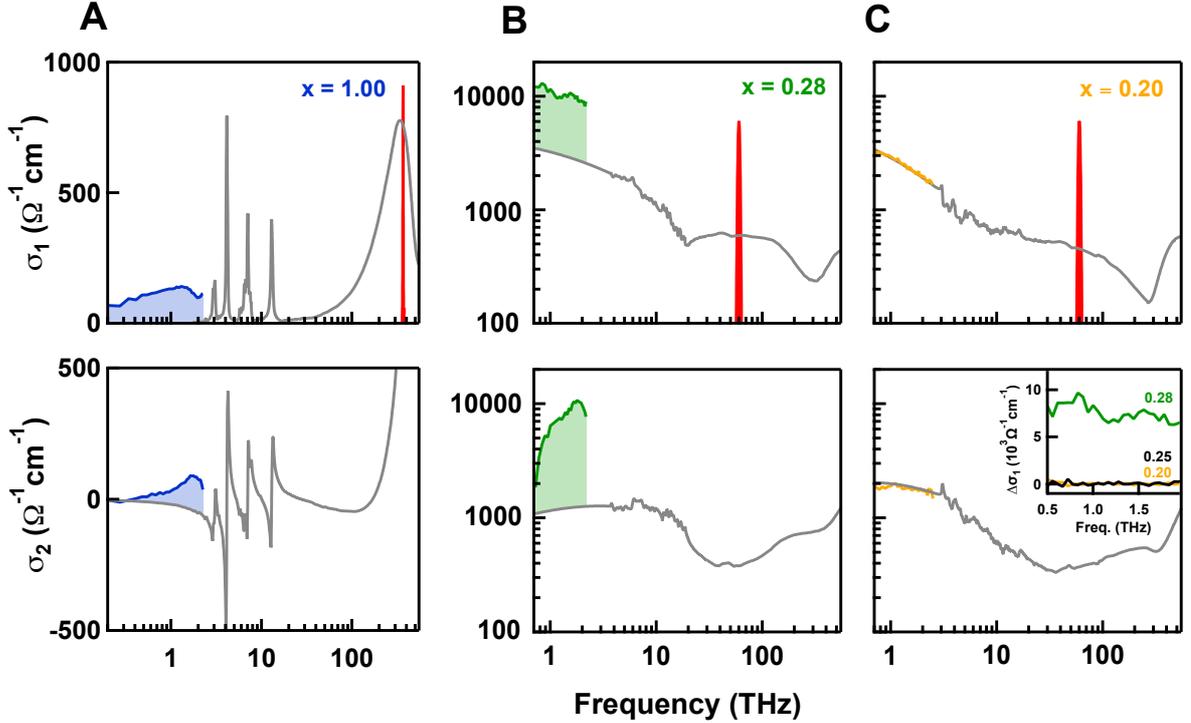

**Figure 2.** Real and imaginary part of the complex conductivity of $BaBiO_3$ (*A*), $BaPb_{0.72}Bi_{0.28}O_3$ (*B*), and $BaPb_{0.8}Bi_{0.2}O_3$ (*C*), at equilibrium (grey) and at τ = 1 ps after photo-excitation (colored), measured at T = 15 K. Steady-state spectra are displayed throughout the whole infrared spectral region, while pump-induced changes (color shaded) were determined only for ω ≲ 2.5 THz. The pump spectrum, tuned either to ~375 THz (*A*) or to ~60 THz (*B*, *C*) to be resonant with the CDW band, is also displayed in red in the top panels. All data have been taken with a pump fluence of ~3.5 mJ/cm². In analogy with $BaPb_{0.8}Bi_{0.2}O_3$ (*C*), also the optimally-doped compound, $BaPb_{0.75}Bi_{0.25}O_3$, displayed no pump-induced response for any T > $T_C$ (see inset with pump-induced conductivity changes).

In a similar fashion, the $x$ = 0.28 sample, belonging to the "insulating" side of the superconducting dome, displayed a dramatic pump-induced conductivity change after photo-stimulation (Fig. 2*B*). Here the Drude peak was already present at equilibrium and experienced an enhancement of a factor of ~3 upon excitation.

On the other hand, both the $x_{opt}$ = 0.25 and $x$ = 0.20 (Fig. 2*C*) compounds yielded no measurable signal at all temperatures T > $T_C$.

Hence, photo-conductivity could only be observed for compositions for which there is evidence for CDW correlations (at least on short range)[42,43]. This may be interpreted in two ways, either as the result of charge excitations across a rigid, semiconducting gap or as a result of (partial) melting of CDW order. In the following analysis, we show that both of these effects are at play, though in different temperature regimes.

In figure 3 we analyze the dynamical evolution of the photo-conductivity of $BaBiO_3$ and $BaPb_{0.72}Bi_{0.28}O_3$ for different temperatures. This photo-conductivity is quantified either by the integrated spectral weight gain (right scale) or by the normalized reflectivity changes (left scale). Both these data analyses show very similar behavior. For both materials, we found a prompt increase of the signal, which reaches its maximum value at $\tau \simeq 1$ ps. At longer time delays, a relaxation dynamics over several picoseconds was observed, characterized by a double-exponential decay (fits are shown as black lines).

The relaxation time constants, $\tau_1$ and $\tau_2$, extracted from the double exponential fits, are displayed in the lower panels as a function of temperature. The reported values ($\tau_1$~0.5 ps and $\tau_2$~10 ps) are consistent with previous experiments on CDW materials[44].

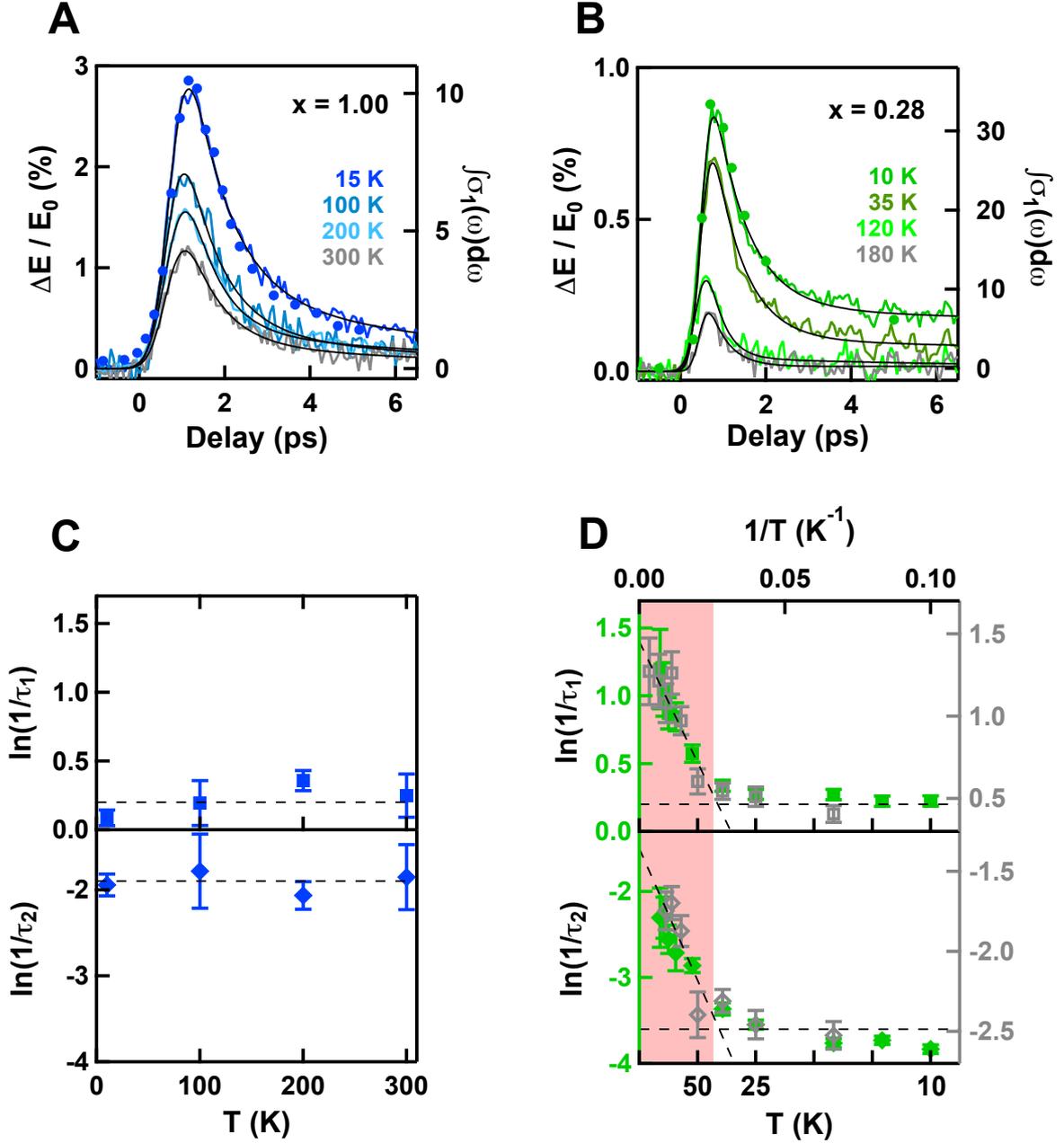

**Figure 3.** Normalized pump-induced changes in the reflected electric field, ΔE/E₀, measured in BaBiO$_3$ (*A*) and BaPb$_{0.72}$Bi$_{0.28}$O$_3$ (*B*) for different temperatures, as a function of pump-probe delay τ. The optical conductivity changes, integrated in the 0.5 – 2 THz range (right scale, shown only at the lowest temperature), follow the same time evolution as ΔE/E₀. In (*C*, *D*) the relaxation time constants τ$_1$ and τ$_2$ are shown, as extracted from the curves in (*A*, *B*) with double exponential fits (black lines). While data in (*C*) are temperature independent, those in (*D*) show a thermally-activated behavior for T ≳ 40 K (red shaded region). Data in (*C*) were taken with the same pump spectrum and fluence as in Fig. 2. Those in (*D*) were partly acquired under the same conditions of Fig. 2 (grey symbols), partly with a detuned pump spectrum (centered at ~25 THz), at a fluence of ~1 mJ/cm² (green symbols).

Importantly, the relaxation dynamics in BaBiO$_3$ (Fig. 3*C*) is almost independent on temperature, which at first glance suggests that the photo-conductivity is likely to result from interband excitations.

On the contrary, in BaPb$_{0.72}$Bi$_{0.28}$O$_3$ (Fig. 3*D*) two well-defined relaxation regimes are found. Below T ≃ 40 K, similar to BaBiO$_3$, no temperature dependence was observed for the relaxation rate. Above T ≃ 40 K, both relaxation time constants showed instead an exponential dependence on temperature, reminiscent of the expected kinetic behavior for a first order phase transition between two distinct thermodynamic phases separated by a free energy barrier[11]. This is quantitatively captured by the slope of the Arrhenius logarithmic plot in Fig. 3*D*, which reflects an activated relaxation of the type exp(−E$_{barrier}$/$k_B$T), with E$_{barrier}$ ≃ 3.5 meV. Finally, the size of the photo-conductivity reduced continuously with temperature, displaying a finite value up to room temperature.

The same experiment was repeated on the *x* = 0.28 compound after detuning the pump spectrum from the center of the CDW band (~25 THz) and reducing the fluence down to ~1 mJ/cm$^2$ (Fig. 3d). Although the size of the photo-conductivity and the τ$_1$ and τ$_2$ values differed from those measured on resonance, a very similar two-regime behavior could be identified in the temperature dependence of both relaxation time constants. Remarkably, also in this case an activated relaxation with E$_{barrier}$ ≃ 3.5 meV was found above T ≃ 40 K, suggesting that these energy scales and temperature dependence of the relaxation rate may be intrinsic to the CDW physics of this material, rather than dependent on the amount of energy deposited into the sample by the laser.

In figure 4 we report a more detailed analysis of the frequency- and time-delay-dependent optical conductivity of both photo-susceptible compounds. We performed

Drude-Lorentz fits on all data, in which the high-frequency Lorentz oscillators were always kept fixed throughout the dynamics, while only the Drude parameters were allowed to vary.

In $BaBiO_3$ (Fig. 4*A*), the measured photo-conductivity could be well fitted by assuming a light-induced *increase* in the carrier plasma frequency at constant scattering rate.

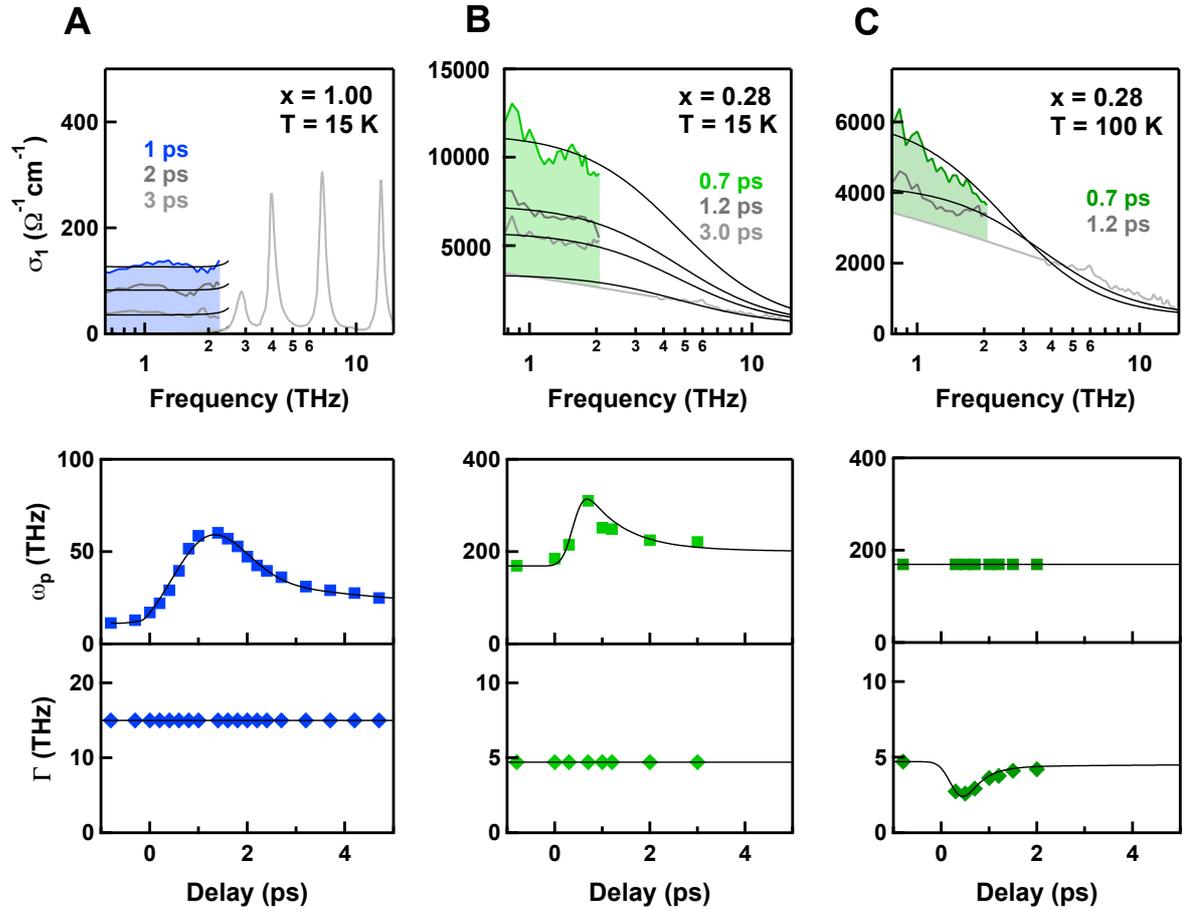

**Figure 4.** Real part of the optical conductivity of $BaBiO_3$ at T = 15 K (*A*), $BaPb_{0.72}Bi_{0.28}O_3$ at T = 15 K (*B*), and $BaPb_{0.72}Bi_{0.28}O_3$ at T = 100 K (*C*), measured at equilibrium (light grey) and at different time delays τ after photo-excitation (colored and dark grey). Drude-Lorentz fits to the data are shown as black lines. Note that a single set of Drude parameters is sufficient to fit all spectra, including those $BaPb_{0.72}Bi_{0.28}O_3$, for which different electronic phases coexist in a dimorphic structure. The extracted time-dependent Drude parameters (plasma frequency, $\omega_p$, and scattering rate, Γ) are reported in the lower panels. All data have been taken with a pump fluence of ~3.5 mJ/cm².

This observation is once again consistent with the response of a semiconductor in which photo-carriers are injected across the gap. Similarly, in the temperature independent regime of photo-conducting $BaPb_{0.72}Bi_{0.28}O_3$ (T = 10 K, see Fig. 4*B*) the enhanced conductivity could be well fitted by an increase in the plasma frequency alone. One may rationalize these observations by assuming that both in the undoped material and in low-temperature $BaPb_{0.72}Bi_{0.28}O_3$ the CDW gap is robust and hence is not perturbed by the photo-excitation.

On the other hand, in the high-temperature regime of $BaPb_{0.72}Bi_{0.28}O_3$ (T = 100 K, see Fig. 4*C*) the data can only be fitted by considering a time independent plasma frequency (carrier density), and rather by assuming a light-induced *decrease* of carrier scattering rate (corresponding to an *increase* in carrier mobility). This effect, taken together with the anomalous Arrhenius-like kinetic behavior (Fig. 3*D*), suggests that photo-conductivity may involve the melting of CDW correlations, which then would recover through nucleation and growth. A 3.5-meV activation barrier can be extracted from the Arrhenius fit of figure 3.

Based on the findings reported above, it appears that for the *x* = 0.28 "underdoped" composition, for temperatures immediately above $T_C$ and up to T ≃ 40 K, the material is characterized by the presence of a robust gap, which, in the absence of long range CDW order must be associated with short range CDW correlations. It is unclear how much of the material (in real space) or the Fermi surface (in momentum space) is associated with this gap[*]. However, such a CDW gap is likely to compete with superconductivity, and hence potentially contributes to the suppression of $T_C$ down to 11 K, compared to the higher values measured in the related compound BKBO (34 K). The nanoscale structural dimorphism reported in Ref. 34 provides a possible means to understand spatial phase separation of CDW and superconducting order, with the former being located in the lower symmetry (orthorhombic) regions and the latter in the high symmetry ones (tetragonal). Such a scenario would explain the apparent granular nature of superconductivity in BPBO for underdoped compositions[34], and the associated superconductor-insulator transition observed in these materials[45].

---

[*] The volume of material that is transiently excited by the pump pulse is significantly larger than the characteristic length scales of the nanoscale structural phase separation[34], and consequently both polymorphs are excited and would contribute to the measured transient response. Remarkably, we can analyse the data of Fig. 4 using a Drude-Lorentz model with only one set of Drude parameters, implying either that CDW correlations derive from just one polymorph (in which case an incomplete volume coverage of the photo-active regions would simply affect the amplitude of the local photo-conductivity) or that similar CDW correlations, characterized by similar parameters, are present in both polymorphs.

Interestingly, our experiments on BaPb$_{0.72}$Bi$_{0.28}$O$_3$ reveal a temperature boundary of ~40 K, which is close to the maximum superconducting transition temperature of potassium-doped bismuthate, as well as to the disorder-free $T_C$ estimated for BPBO in Ref. 35. In addition, the same energy scale (3.5 meV $\simeq$ 40 K) is also found here to govern the relaxation kinetics of the photo-conductivity of the 28%-doped compound at high temperatures.

We speculate that for T $\gtrsim$ 40 K, where the photo-conductivity is dominated by changes in mobility, a "soft" CDW gap may be readily melted by the photo-excitation, presumably generating a transient metallic phase. One can then imagine that relaxation back to the ground state may involve barrier crossing, nucleation and growth. Precise details of such a mechanism await development of a clearer microscopic picture of the ways in which the competing electronic phases coexist in the context of the dimorphic structure.

In summary, we have reported the transient THz conductivity of BaPb$_{1-x}$Bi$_x$O$_3$ at different doping levels and temperatures after photo-excitation across the charge-density-wave band. In BaBiO$_3$ and in the low-temperature regime of BaPb$_{0.72}$Bi$_{0.28}$O$_3$ we measured a temperature-independent relaxation dynamics, consistent with a scenario of transient photo-carriers pumped across the semiconducting gap of a rigid CDW. For the doped compound this observation adds weight to previous arguments suggesting the existence of short range CDW correlations in this part of the phase diagram. A qualitatively different behavior was observed instead in BaPb$_{0.72}$Bi$_{0.28}$O$_3$ at higher temperatures (T $\gtrsim$ 40 K), where the photo-conductivity relaxation showed the expected kinetic behavior for a transition between two distinct thermodynamic phases separated by a free energy barrier of 3.5 meV $\simeq$ 40 K. This universal energy scale, which regulates the relaxation kinetics of underdoped BPBO, is surprisingly close to the superconducting transition temperature of the related Ba$_{1-x}$K$_x$BiO$_3$ compound, an observation which opens up new perspectives on the nature of the interplay between superconductivity and CDW formation in the bismuthate superconductors.

**Acknowledgments**

The research leading to these results received funding from the European Research Council under the European Union's Seventh Framework Programme (FP7/2007-2013)/ERC Grant Agreement No. 319286 (QMAC). We acknowledge support from the Deutsche Forschungsgemeinschaft via the excellence cluster 'The Hamburg Centre for Ultrafast Imaging—Structure, Dynamics and Control of Matter at the Atomic Scale' and the priority program SFB925. Work at Stanford University was supported by the AFOSR, Grant No. FA9550-09-1-0583.

**REFERENCES**

[1] Si Q, Yu R, Abrahams E (2016) High-temperature superconductivity in iron pnictides and chalcogenides. *Nature Rev. Mater.* 1:16017.

[2] Tranquada JM, Sternlieb BJ, Axe JD, Nakamura Y, Uchida S (1995) Evidence for stripe correlations of spins and holes in copper oxide superconductors. *Nature* 375:561-563.

[3] Ghiringhelli G, Le Tacon M, Minola M, Blanco-Canosa S, Mazzoli C, Brookes NB, De Luca GM, Frano A, Hawthorn DG, He F, Loew T, Moretti Sala M, Peets DC, Salluzzo M, Schierle E, Sutarto R, Sawatzky GA, Weschke E, Keimer B, Braicovich L (2012) Long-Range Incommensurate Charge Fluctuations in (Y,Nd)Ba$_2$Cu$_3$O$_{6+x}$. *Science* 337:821-825.

[4] Chang J, Blackburn E, Holmes AT, Christensen NB, Larsen J, Mesot J, Ruixing Liang, Bonn DA, Hardy WN, Watenphul A, von Zimmermann M, Forgan EM, Hayden SM (2012) Direct observation of competition between superconductivity and charge density wave order in YBa$_2$Cu$_3$O$_{6.67}$. *Nat. Phys.* 8, 871-876.

[5] Comin R, Frano A, Yee MM, Yoshida Y, Eisaki H, Schierle E, Weschke E, Sutarto R, He F, Soumyanarayanan A, He Y, Le Tacon M, Elfimov IS, Hoffman JE, Sawatzky GA, Keimer B, Damascelli A (2014) Charge Order Driven by Fermi-Arc Instability in Bi$_2$Sr$_{2-x}$La$_x$CuO$_{6+\delta}$. *Science* 343:390-392.

[6] da Silva Neto EH, Aynajian P, Frano A, Comin R, Schierle E, Weschke E, Gyenis A, Wen J, Schneeloch J, Xu Z, Ono S, Gu GD, Le Tacon M, Yazdani A (2014) Ubiquitous Interplay Between Charge Ordering and High-Temperature Superconductivity in Cuprates. *Science* 343:393-396.

[7] Hücker M, von Zimmermann M, Debessai M, Schilling JS, Tranquada JM, Gu GD (2010) Spontaneous Symmetry Breaking by Charge Stripes in the High Pressure Phase of Superconducting La$_{1.875}$Ba$_{0.125}$CuO$_4$. *Phys. Rev. Lett.* 104:057004.

[8] Hücker M, von Zimmermann M, Xu ZJ, Wen JS, Gu GD, Tranquada JM (2013) Enhanced charge stripe order of superconducting La$_{2-x}$Ba$_x$CuO$_4$ in a magnetic field. *Phys. Rev. B* 87:014501.

[9] Nicoletti D, Cavalleri A (2016) Nonlinear light–matter interaction at terahertz frequencies. *Adv. Opt. Photon.* 8:401-464.

[10] Fausti D, Tobey RI, Dean N, Kaiser S, Dienst A, Hoffmann MC, Pyon S, Takayama T, Takagi H, Cavalleri A (2011) Light-Induced Superconductivity in a Stripe-Ordered Cuprate. *Science* 331:189-191.

[11] Hunt CR, Nicoletti D, Kaiser S, Takayama T, Takagi H, Cavalleri A (2015) Two distinct kinetic regimes for the relaxation of light-induced superconductivity in La$_{1.675}$Eu$_{0.2}$Sr$_{0.125}$CuO$_4$. *Phys. Rev. B* 91:020505(R).

[12] Nicoletti D, Casandruc E, Laplace Y, Khanna V, Hunt CR, Kaiser S, Dhesi SS, Gu GD, Hill JP, Cavalleri A (2014) Optically-induced superconductivity in striped $La_{2-x}Ba_xCuO_4$ by polarization-selective excitation in the near infrared. *Phys. Rev. B* 90:100503(R).

[13] Casandruc E, Nicoletti D, Rajasekaran S, Laplace Y, Khanna V, Gu GD, Hill JP, Cavalleri A (2015) Wavelength-dependent optical enhancement of superconducting interlayer coupling in $La_{1.885}Ba_{0.115}CuO_4$. *Phys. Rev. B* 91:174502.

[14] Khanna V, Mankowsky R, Petrich M, Bromberger H, Cavill SA, Möhr-Vorobeva E, Nicoletti D, Laplace Y, Gu GD, Hill JP, Först M, Cavalleri A, Dhesi SS (2016) Restoring interlayer Josephson coupling in $La_{1.885}Ba_{0.115}CuO_4$ by charge transfer melting of stripe order. *Phys. Rev. B* 93:224522.

[15] Först M, Tobey RI, Bromberger H, Wilkins SB, Khanna V, Caviglia AD, Chuang YD, Lee WS, Schlotter WF, Turner JJ, Minitti MP, Krupin O, Xu ZJ, Wen JS, Gu GD, Dhesi SS, Cavalleri A, Hill JP (2014) Melting of Charge Stripes in Vibrationally Driven $La_{1.875}Ba_{0.125}CuO_4$: Assessing the Respective Roles of Electronic and Lattice Order in Frustrated Superconductors. *Phys. Rev. Lett.* 112:157002.

[16] Hu W, Kaiser S, Nicoletti D, Hunt CR, Gierz I, Hoffmann MC, Le Tacon M, Loew T, Keimer B, Cavalleri A (2014) Optically enhanced coherent transport in $YBa_2Cu_3O_{6.5}$ by ultrafast redistribution of interlayer coupling. *Nat. Mater.* 13:705–711.

[17] Kaiser S, Hunt CR, Nicoletti D, Hu W, Gierz I, Liu HY, Le Tacon M, Loew T, Haug D, Keimer B, Cavalleri A (2014) Optically induced coherent transport far above $T_c$ in underdoped $YBa_2Cu_3O_{6+\delta}$. *Phys. Rev. B* 89:184516.

[18] Hunt CR, Nicoletti D, Kaiser S, Pröpper D, Loew T, Porras J, Keimer B, Cavalleri A (2016) Dynamical decoherence of the light induced inter layer coupling in $YBa_2Cu_3O_{6+\delta}$. *Phys. Rev. B* 94:224303.

[19] Först M, Frano A, Kaiser S, Mankowsky R, Hunt CR, Turner JJ, Dakovski GL, Minitti MP, Robinson J, Loew T, Le Tacon M, Keimer B, Hill JP, Cavalleri A, Dhesi SS (2014) Femtosecond x rays link melting of charge-density wave correlations and light-enhanced coherent transport in $YBa_2Cu_3O_{6.6}$. *Phys. Rev. B* 90:184514.

[20] Pickett WE (2001) The other high-temperature superconductors. *Physica B* 296:112-119.

[21] Sleight AW, Gillson JL, Bierstedt PE (1975) High-temperature superconductivity in the $BaPb_{1-x}Bi_xO_3$ systems. *Solid State Commun.* 17:27-28.

[22] Cava RJ, Batlogg B, Krajewski JJ, Farrow R, Rupp Jr. LW, White AE, Short K, Peck WF, Kometani T (1988) Superconductivity near 30 K without copper: the $Ba_{0.6}K_{0.4}BiO_3$ perovskite. *Nature* 332:814-816.

[23] Sleight AW (2015) Bismuthates: $BaBiO_3$ and related superconducting phases. *Physica C* 514:152-165.

[24] Uchida S, Kitazawa K, Tanaka S (1987) Superconductivity and metal-semiconductor transition in $BaPb_{1-x}Bi_xO_3$. *Phase Transit.* 8:95-128.

[25] Taraphder A, Pandit R, Krishnamurthy HR, Ramakrishnan TV (1996) The exotic barium bismuthates. *Int. J. Mod. Phys. B* 10:863.

[26] Cox DE, Sleight AW (1976) Crystal structure of $Ba_2Bi^{3+}Bi^{5+}O_6$. *Solid State Commun.* 19:969-973.

[27] Pei S, Jorgensen JD, Dabrowski B, Hinks DG, Richards DR, Mitchell AW, Newsam JM, Sinha SK, Vaknin D, Jacobson AJ (1990) Structural phase diagram of the $Ba_{1-x}K_xBiO$ system. *Phys. Rev. B* 41:4126-4141.

[28] Marx DT, Radaelli PG, Jorgensen JD, Hitterman RL, Hinks DG, Pei S, Dabrowski B (1992) Metastable behavior of the superconducting phase in the $BaBi_{1-x}Pb_xO_3$ system. *Phys. Rev. B* 46:1144-1156.

[29] Tajima S, Yoshida M, Koshizuka N, Sato H, Uchida S (1992) Raman-scattering study of the metal-insulator transition in $Ba_{1-x}K_xBiO_3$. *Phys. Rev. B* 46:1232-1235.

[30] Yacoby Y, Heald SM, Stern EA (1997) Local oxygen octahedral rotations in $Ba_{1-x}K_xBiO_3$ and $BaBiO_3$. *Solid State Commun.* 101:801-806.

[31] Battlog B, Cava RJ, Rupp Jr. LW, Mujsce AM, Krajewski JJ, Remeika JP, Peck Jr. WF, Cooper AS, Espinosa GP (1998) Density of States and Isotope Effect in BiO Superconductors: Evidence for Nonphonon Mechanism. *Phys. Rev. Lett.* 61:1670-1673.

[32] Yin ZP, Kutepov A, Kotliar G (2013) Correlation-Enhanced Electron-Phonon Coupling: Applications of GW and Screened Hybrid Functional to Bismuthates, Chloronitrides, and Other High-Tc Superconductors. *Phys. Rev. X* 3:021011.

[33] Climent-Pascual E, Ni N, Jia S, Huang Q, Cava RJ (2011) Polymorphism in $BaPb_{1-x}Bi_xO_3$ at the superconducting composition. *Phys. Rev. B* 83:174512.

[34] Giraldo-Gallo P, Zhang Y, Parra C, Manoharan HC, Beasley MR, Geballe TH, Kramer MJ, Fisher IR (2015) Stripe-like nanoscale structural phase separation in superconducting $BaPb_{1-x}Bi_xO_3$. *Nat. Comms.* 6:8231.

[35] Luna K, Giraldo-Gallo P, Geballe T, Fisher IR, Beasley M (2014) Disorder Driven Metal Insulator Transition in $BaPb_{1-x}Bi_xO3$ and Inference of Disorder-Free Critical Temperature. *Phys. Rev. Lett.* 113:177004.

[36] Giraldo-Gallo P, Lee H, Beasley MR, Geballe TH, Fisher IR (2013) Inhomogeneous Superconductivity in $BaPb_{1-x}Bi_xO3$. *J. Supercond. Nov. Magn.* 26:2675-2678.

[37] Tajima S, Uchida S, Masaki A, Takagi H, Kitazawa K, Tanaka S (1985) Optical study of the metal-semiconductor transition in $BaPb_{1-x}Bi_xO_3$. *Phys. Rev. B* 32:6302-6311.

[38] Tajima S, Uchida S, Masaki A, Takagi H, Kitazawa K, Tanaka S, Sugai S (1987) Electronic states of BaPb$_{1-x}$Bi$_x$O$_3$ in the semiconducting phase investigated by optical measurements. *Phys. Rev. B* 35:696-703.

[39] Puchkov A, Timusk T, Karlow MA, Cooper SL, Han PD, Payne DA (1996) Doping dependence of the optical properties of Ba$_{1-x}$K$_x$BiO$_3$. *Phys. Rev. B* 54:6686-6692.

[40] Schlesinger Z, Collins RT, Scott BA, Calise JA (1998) Superconducting energy gap of BaPb$_{1-x}$Bi$_x$O$_3$. *Phys. Rev. B* 38:9284-9286(R).

[41] Dressel M, Grüner G (2002) *Electrodynamics of Solids*, Cambridge University Press, Cambridge.

[42] Boyce JB, Bridges FG, Claeson T, Geballe TH, Li GG, Sleight AW (1991) Local structure of BaPb$_{1-x}$Bi$_x$O$_3$ determined by x-ray-absorption spectroscopy. *Phys. Rev. B* 44:6961–6972.

[43] Menushenkov AP (1998) On the mechanism of superconductivity in BaBi(Pb)O$_3$-Ba(K)BiO$_3$ systems: analysis of XAFS-spectroscopy data. *Nucl. Instr. and Meth. in Phys. Res. A* 405:365–369.

[44] Demsar J, Biljakovic K, Mihailovic D (1999) Single Particle and Collective Excitations in the One-Dimensional Charge Density Wave Solid K$_{0.3}$MoO$_3$ Probed in Real Time by Femtosecond Spectroscopy. *Phys. Rev. Lett.* 83:800-803.

[45] Giraldo-Gallo P, Lee H, Zhang Y, Kramer MJ, Beasley MR, Geballe TH, Fisher IR (2012) Field-tuned superconductor-insulator transition in BaPb$_{1-x}$Bi$_x$O$_3$. *Phys. Rev. B* 85:174503.